\documentclass[pdftexify,conference]{IEEEtran}
\usepackage{cite}
\usepackage{amsmath, amssymb, amsfonts, bbm, amsthm, eqparbox}
\usepackage{array, float}
\usepackage{graphicx,epsfig,subfig}
\usepackage[colorlinks,linkcolor=black,urlcolor=black,citecolor=black]{hyperref}
\usepackage{url}
\usepackage[latin1]{inputenc}
\usepackage[T1]{fontenc}
\IEEEoverridecommandlockouts

\newcommand{\MYHeader}{IV Workshop-school on Quantum Computation and Information (WECIQ 2012)}

\makeatletter

\def\ps@headings{%
\def\@oddhead{\MYHeader \hfil}
\def\@evenhead{\MYHeader \hfil }
\def\@oddfoot{\ }%
\def\@evenfoot{\ }}

\def\ps@IEEEtitlepagestyle{%
\def\@oddhead{\MYHeader \hfil}
\def\@evenhead{\MYHeader \hfil}
\def\@oddfoot{\ }%
\def\@evenfoot{\ }}

\makeatother

\begin{document}

\title{Quantum Memories via Electromagnetically Induced Transparency in Three-Level Systems}

\author{F. Revson Fernandes Pereira, Danieverton Moretti, Elloá B. Guedes
\thanks{This work was supported in part by the CNPq and CAPES. F. Revson Fernandes Pereira and Danieverton Moretti are with the Physics Department at the Federal University of Campina Grande. Elloá B. Guedes is with the Institute for Studies in Quantum Computation and Information (IQuanta) also at the Federal University of Campina Grande. R. Aprígio Veloso, 882. Campina Grande, Paraíba, Brasil. E-mails: \texttt{revson.ee@gmail.com}, \texttt{dmoretti@df.ufcg.edu.br}, \texttt{elloaguedes@gmail.com}.}}

\maketitle

\begin{abstract}
This work aims at analyzing the adequacy of Laguerre-Gauss (LG) beams and three-level quantum systems to build quantum memories. We focus on such systems in which there is a phenomenon called Eletromagnetic Induced Transparency (EIT) which enables the information storage in atomic cells. We show both theoretical and practical results regarding LG beams and the conditions according to which EIT raises. The existence of such conditions is necessary to the implementation of quantum memories in this scenario.
\end{abstract}


\section{Main Results}

This work was performed in two parts: theoretical and practical ones. In the theoretical part, we studied the Laguerre-Gauss (LG) beams construction, characterization and properties (linear and angular momenta, Poynting vector, and other modes) and the conditions to the existence of Eletromagnetic Induced Trasparency (EIT) in the three-level systems, including its mathematical model and also its applications.

\begin{figure}[H]
  \centering
  \includegraphics[width=0.8\linewidth]{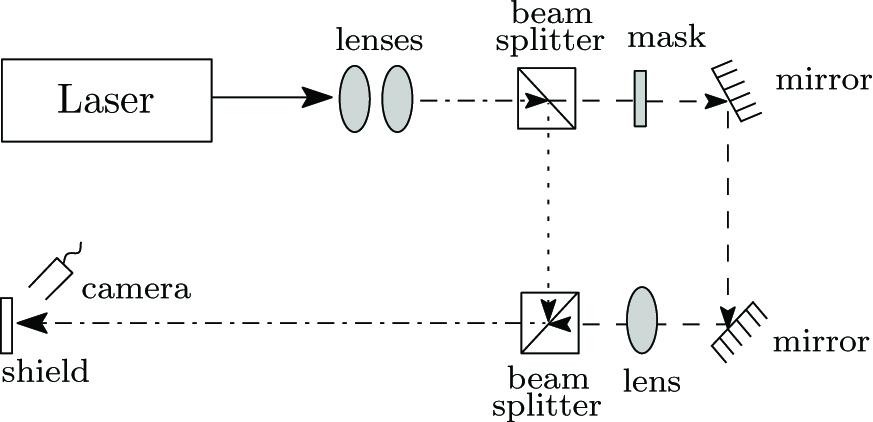}\\
  \caption{Experimental apparatus to the preparation of LG modes.}\label{fig:esquema}
\end{figure}

Regarding the practical part, we implemented the holographic method to produce LG beams with and without phase singularity. To do so, we set up the apparatus showed in Figure \ref{fig:esquema}, which was used to produce LG with modes $\ell = 0$ and $p=0$, $\ell = 1$ and $p = 0$, and $\ell = 2$ and $p = 0$. The terms $\ell$ and $p$ denote the azimuthal and the radial indexes of LG beams.

\begin{figure}[H]
\centering
\subfloat[]{\label{fig:modo0}\includegraphics[width=0.15 \textwidth]{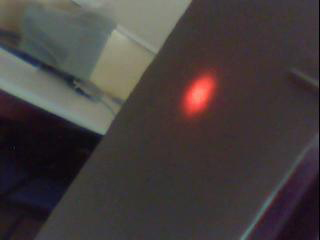}} \hspace{0.2cm}
\subfloat[]{\label{fig:modo1}\includegraphics[width=0.15 \textwidth]{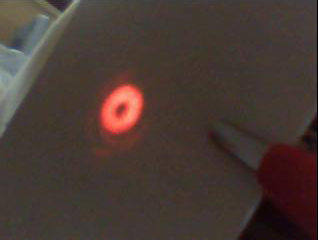}}
\hspace{0.2cm}
\subfloat[]{\label{fig:modo2}\includegraphics[width=0.15 \textwidth]{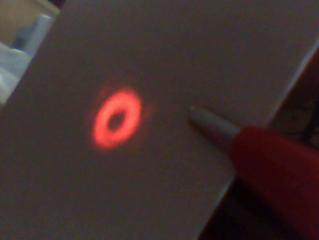}}
\caption{Images obtained from the camera from the apparatus described on Figure \ref{fig:esquema}.}
\label{fig:modos}
\end{figure}

The results obtained using the described apparatus are shown on Figure \ref{fig:modos}. Each Figure \ref{fig:modo0}, \ref{fig:modo1}, and \ref{fig:modo2} correspond to a configuration  $\ell = 0$ and $p=0$, $\ell = 1$ and $p = 0$, and $\ell = 2$ and $p = 0$, respectively.

To the results illustrated in Figures \ref{fig:modo1} and \ref{fig:modo2}, the experimental error was of $6\%$ and $8\%$, respectively. It shows confidence in the results obtained.

The LG beams mentioned were used in a three-level system to produce
the EIT phenomenon. In such systems there are two Zeeman ground
states ($\left| b\right\rangle$ and $\left| c\right\rangle$) and one excited state ($\left| a\right\rangle$).
Under the action of beams with angular momentum, the EIT occurs
which consists in the decrease of light absorption thanks to an
atomic ensemble when two fields are, or are not, in resonance with
two different transitions ($\left| a\right\rangle \leftrightarrow \left| b\right\rangle$ and
$\left| a\right\rangle \leftrightarrow \left| c\right\rangle$), having one state in common
($\left|a\right\rangle$).

\begin{figure}[H]
  \centering
  \includegraphics[width=0.85\linewidth]{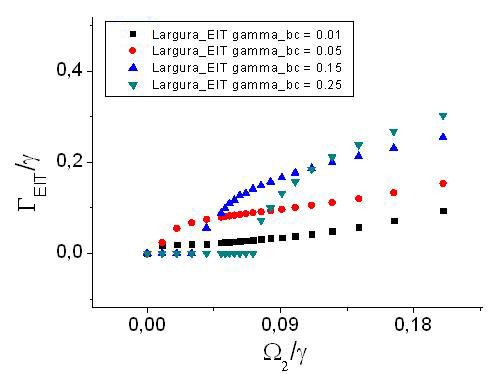}\\
  \caption{Simulation results.}\label{fig:largura}
\end{figure}

A simulation performed aiming at measure the width on the average height of the EIT phenomenon produced the results shown in Figure \ref{fig:largura}. These results considering $\gamma_{bc}$ equal to $0,15 \gamma$ and $0,25 \gamma$ are particularly interesting because they are equal to $0$ considering low values of field intensity of the control beam. Quantum memories based on EIT demand the occurrence of such values. So, to make use of quantum memories using EIT, such EIT characteristics must be preserved.

The next steps of our research consist in the implementation of the simulated scenario and in the practical verification of the phenomenon occurred in such quantum memories.
\end{document}